\documentstyle[12pt,version2,aps]{revtex}
\begin{document}

\vfill
\eject

\renewcommand{\theequation}{\arabic{section}.\arabic{equation}}
\newcommand{\eqreset}{\setcounter{equation}{0}}
\setlength{\textheight}{20cm}
\vspace*{.9 in}
\begin{center}
{\large\bf                    FLUCTUATIVE MECHANISM OF VORTEX
                    NUCLEATION IN THE FLOW OF $^4He$.$^*$}

\vspace{.5 in}
{\sc F.V. Kusmartsev}

 \vspace{.3 in}

{\it L.D. Landau Institute for Theoretical Physics, 
Moscow, Russia \par 
 Dep. of  Phys., Abo Akademy, SF-20500, Abo, Finland$^*$
}
and\\
\ \\
 { \it NORDITA, Blegdamsvej, 17, Kopenhagen 0, Danmark \\
e:mail:     Feo@NORDITA.DK}

\end{center}

 \vspace{.9 in}

\begin{abstract}

We propose a mechanism of a vortex nucleation
in a flow of a rotating superfluid $^4He$.  The mechanism
is related to the  creation by critical fluctuations
of  a "plasma" of
 half-vortex rings  located
 near the wall.
The  "plasma" screens the attraction of the vortex to the wall and
permits   vortex nucleation.
 In the spirit of Williams-Shenoy
theory we derive the scaling laws in the critical region and estimate
the scaling relation and the critical exponent $p$
 for critical velocity; we find  $V_c\sim V_{0}(1-T/T_c)$, so
that $p=1$. Various applications of the obtained results are discussed.

 \end{abstract}
 
\vfill
\eject

    The studies of the vortex nucleation in the flow $^4He$ have a long
history and many interesting results. Packard and Sanders \cite{Pack}
first found that the nucleation of the vortices may be related with
the rotation of the $^4He$. It is, however, recognized that there is
a very large barrier near the wall, which prevents the free vortices
from penetrating into the volume. On the other hand Awshalom and Schwarz\cite{Awsh} have shown the presence 
of remnant
vortices, pinned by the wall.  It
was established that the vortex nucleation in a submicron orifice is related to the activation or tunneling of the half-vortex ring    into the volume
of the superfluid and  different mechanisms  have been proposed\cite{Ihas,Davi,Varoq,Ihas1}. 

However, it was, experimentally,
found that the critical velocity has the scaling law of the type
$V_c=V_{0} (1-T/T_0)$
\cite{Varo,Ihas,Davi,Varoq,Ihas1,Tyle} or a small
deviation from this law \cite{Amar,Bona}. 
For the very small orifice, the barrier height for  vortex nucleation
is small. Therefore, activational and tunneling processes for the single vortex
generation are possible. However, when  the orifice radius increases the barrier
height increases. The larger barrier height occurs  for penetration of
vortices in rotating superfluid\cite{Pack}. Therefore in these cases both the activational and the
tunneling mechanisms alone for the vortex nucleation are ruled out.
 To improve
a disagreement in the vortex nucleation rate between  experiment and 
Iordanskii-Langer-Fisher (ILF) nucleation theory\cite{Iord,Lang},
 Kawatra, Pathria\cite{Kawa} and Volovik \cite{Volo} 
suggested that near the wall  there is a microscopical surface vorticity sheet.
More emphases to the surface vorticity sheet  has been assigned in next
development by Ihas et al \cite{Ihas,Varoq,Ihas1} who suggested that 
 the surface vorticity sheet   may 
aid to tunneling or   activation 
through a  velocity dependent barrier.

 We   propose a mechanism, which  is very different from
the one discussed in the literature\cite{Ihas,Davi,Varoq,Ihas1,Tyle}.
 The vortex penetrates the barrier near the wall with the aid of
 critical fluctuations via a creation of a half-vortex ring  "plasma"( or the surface vorticity sheet )  .
There occurs
a phase transition, in which 
{\it the  width of microscopical the surface vorticity sheet  reaches a critical size $R_c$},
the barrier for the vortex nucleation  disappears and  vortex generation
 is started.   The vortex nucleates in a process similar to the
Berezinskii-Kosterlitz-Thouless(BKT) phase transition 
\cite{Berezinski,KT1,KT2}, where instead of
  vortex-antivortex (V-A) pairs there near the wall   many  half-vortex rings  (fluctuational plasma or the surface vorticity sheet   are generated.
 The vortex coupling with the wall
 is screened and the half-vortex rings  of a very large radius
are  created. The longer half-ring, then, nucleates the vortex
and initiates phase slippage.   With the increase of the flow velocity
and the temperature, the correlation length of critical
fluctuations in the surface vorticity sheet or its thickness   increases.
This growth is stopped at some threshold defined both by the temperature and by the flow velocity where the 
creation of vortices is started. Sonin \cite{SingleRing,Soni}   noticed that
 "neither tunneling nor an activation
is a threshold effect, but the vortex nucleation  is a threshold effect".
 Our mechanism  describes indeed the threshold effect.

 As the flow velocity
increases the energy of the pinned the half-vortex ring  decreases.
This stimulates
their activation through thermal fluctuations.
 In their turn the
fluctuational half-vortex rings  of a small radius  assist
in the creation of
 half-vortex rings  of larger radius and so on.
The picture is reminiscent of
the scaling in BKT transition, where 
 the coupling between the vortex and
antivortex  decreases as the temperature rises.
 The similar situation occurs in the Williams-Shenoy(WS)
\cite {Will,Shen} model of the $\lambda-$phase transition where the role of V-A pairs is
played by  the  vortex rings.
  The role of half-vortex rings  in our mechanism are similar to the role
of the  vortex-antivortex  pairs,
whose spontaneous generation is a driving mechanism of the
BKT transition.

We  derive 
scaling relations associated with two relevant operators:
 the temperature and
the flow velocity. In this derivation we will follow
 the WS approach \cite {Will,Shen}. 
  We start with an assumption that the half-vortex rings  are pinned by the wall and are  polarized by the flow. To create a half-vortex
ring in the external flow of the velocity $u_0$ one needs the energy

\begin{equation}
    E= E_0-{ p u_0 \cos \theta}
\end{equation}
where $E_0$, $p$ are the energy and the impulse of the half-vortex ring  .
The angle between the flow velocity ${\bf u_0}$
and the normal to the vortex ring plane is $\theta$.
The energy is equal to the half of the energy  of the
vortex ring \cite{Fetter76}
\begin{equation}
E_0 = \pi^2 R K_0(\ln \frac{R}{a_c} + C)
\end{equation}
where $R$ is the radius of the loop; $C$ and $a_c$ are core energy and
radius, respectively. The constant $K_0$ is   proportional to superfluid density.  The half- impulse is equal to $    p = {k\pi R^2}/{2}$,
where $k$ is a vorticity. The probability to find a single
half-vortex ring   on a scale $R$ in the area 
 $ \pi R^2 dR d\cos\theta $ is
 $dn(R,\theta) = \pi R^2dR d\cos\theta \exp(-E(R,\theta))$. 

    In the low fugacity limit\cite{Will}
the interaction between vortex rings is neglected. The effective susceptibility is equal to $\chi = \int_1^{R_c} dn(R,\theta) \alpha_{\Delta}$, where
$\alpha_{\Delta}$ is a polarizability of these half-vortex rings ,
which is estimated to be a quarter of the polarizability of a vortex
ring $ \alpha_{\Delta}=\beta p^2/12$ .  The scale is measured in units
of $a_c$.  Following Williams and Shenoy one may introduce the dielectric
constant $\epsilon = 1+4\pi\chi$ and the screened density as
$K_r={K_0}/{\epsilon}$. As the result we arrive to the recursion equation:
\begin{equation}
\frac{1}{K_r} = \frac{1}{K} + (A/2) \int_1^{R_c} dR d\cos \theta R^6 \exp(-E(R,\theta))
\label{rek}
\end{equation}
where $A$ is a some constant. Following Ref.\cite{Will} we introduce the fugacity as $y=\exp[-\pi^2 K \ln\sqrt{gK} +C]$; then after an angular
integration the recursion
relation takes the general   BKT-WS form:
\begin{equation}
\frac{1}{K_r} = \frac{1}{K} + A \int_1^{R_c} dR 
                R^4 y^R \exp(-\pi^2 K R \ln R ) \sinh (u R^2)/u
\label{main-rek},
\end{equation}
where we   introduced the notations as $u=k \pi u_0/2$.
    With the aid of the vortex core rescaling technique by Jose et al \cite{Jose} which consists of integrating over small distances 
and then rescaling as $R\rightarrow R b$, we get the following relation:
\begin{equation}
\frac{1}{K_r} = \frac{1}{K} + A y \ln b \frac{\sinh u}{u}  + A b^5 \int_1^{R_c} dR R^4 y^{bR} \exp(-\pi^2 K bR \ln bR)  \frac{\sinh (u b^2 R^2)}{u}
\label{resume-rek}
\end{equation}

By introducing the new variables
\begin{eqnarray}
\frac{1}{{K}^\prime} = \frac{1}{b} (\frac{1}{K} + A y \ln b \frac{\sinh u}{u} ) \\
{A}^\prime {y}^\prime =A b^6 y^b \exp({-\pi^2 K b \ln b}) 
\label{renorm-eq1}
\end{eqnarray}
as well as ${K_r}^\prime = b K_r$ and ${u}^\prime=b^2 u$ we get the same
relation as eq(\ref{main-rek}). These rescaling eqs may
be represented in differential form similar to that of Williams \cite{Will} 
\begin{eqnarray}
    \frac{d}{dl}(\frac{1}{K}) = -\frac{1}{K} + A_0 Ky\frac{\sinh u}{u} \\
    \frac{dy}{dl} = [6 - K(\ln\sqrt{Kg}+C+1)]y
\end{eqnarray}
with the initial value $g_{0}=1/K_0$ or  
   in the form   similar to that of
 Shenoy for $3D XY$ model:
\begin{eqnarray}  
    \frac{dK}{dl} = K - A_0 y K^2 \frac{\sinh u}{u} \\
    \frac{dy}{dl} = (6- \pi^2 KL) y\\
    \frac{du}{dl} = 2u
\label{renorm-eq2}
\end{eqnarray}
where $L= \ln\frac{a}{a_c} + 1$ and $a,a_c$ are an effective size of the loop and  of its core, respectively(above $a_c=1$)\cite{Shen}.
 One sees that the difference with Williams-Shenoy 
 equations lies in the coefficient $A_0$ and in 
the strong (exponential) dependence on the flow velocity, which appears
in the second term of the  first equations.
 In other words when $u=0$ the equations coincide with
WS equations, but with the difference that Shenoy got the value $A_0=\frac{4\pi^3}{3}$, while Williams --- the value $A_0=\frac{\pi^5}{6}$.
Because of the wall, the half-rings are created on a half space, so
the coefficient in our equations is equal to 1/8 of the Williams one, $A_0=\pi^5/48$.
The dependence of scaling on the superfluid flow velocity $u$
arises because of the spontaneous generation
of half-vortex rings  induced by the flow.
This diminishes the effective coupling $K$ and stimulates
  the vortex nucleation. One sees from this equation that the flow velocity
is very important for the behavior of $K$ \cite{Rem2}.

There are two types of a behavior of the system.
 The first one is the superfluid or the low temperature one, which is characterized by
growing $u_l=u_0 e^{2l}$, $K_l=K_0 e^l$ and vanishing fugacity 
$y=y_0 e^{-\frac{a}{\xi_0}}$, where $\xi_0=\frac{1}{(\pi^2K_0L)}$ and
valid when $A_0 K y \frac{\sinh u}{u} \ll 1$. The second one is a high-temperature
solution, which is characterized by growing the fugacity $y$.
One sees that the transition of the vortex nucleation
depends explicitly from the flow velocity, while the transition
temperature does not.

    Between these two low- and high-temperature phases there is a critical point, which is associated with the nontrivial fixed point of the rescaling equations: $  u = 0$,
\begin{eqnarray}
      (6-\pi^2 K L) y = 0\\
    K - A_0 y K^2 \frac{\sinh u}{u}  = 0
\end{eqnarray}
The nontrivial solution is 
$
K_1 = \frac{6}{\pi^2L}
$ and
$
y_1 = \frac{\pi^2L}{6A_0}.
$

Now let us make an expansion in the vicinity of this critical point as
$K_l = K_1(1+k), y_l = y_1(1+y)$, and $u_l = u$. Then, the scaling
equations take the linear form:
\begin{eqnarray} 
  \frac{du}{dl} = 2u\\
   \frac{dy}{dl} = -6k\\
   \frac{dk}{dl} = -k-y
\end{eqnarray}

 The behavior of scaling near the fixed point is
associated with the three eigenvalues: $\lambda_{+} =2$,
 $\lambda_{-} =-3$ and
$\lambda_3=2$. 
The rescaling law for the free energy  $F_l$  obeys the relation
\begin{equation}
Z(K_0,y_0,u_0) = e^{-(F_l-F_0)L^3} Z(K_l,y_l,u_l),
\end{equation}
which is associated with two relevant operators
related to the temperature axis
 $A_+ e^{\lambda_+l}$ and the critical velocity $u_+ e^{\lambda_3 l}$ and one
irrelevant associated with the fugacity $y_l$, i.e.
\begin{equation}
Z(K_0,y_0,u_0) = e^{-(F_l-F_0)L^3} Z(A\mid\epsilon\mid e^{\lambda_+l}, A_- e^{\lambda_-l}, u_+ e^{\lambda_3 l})
\end{equation}
where $\epsilon$ is a deviation of the temperature $T$ from $T_c$:
 $\epsilon=(1-T/T_c)$ and $u_+$ is the ratio of critical velocity to
the relative one $u_+= V_c/V_0$\cite{Gill}.
Because of the two relevant operators it is nontrivial to find where
the scaling  must be stopped. In order to understand this we must look into the original recursion relation (\ref{rek}). 
With the scaling the critical velocity $u_l$ and the coherence length $\xi$ are growing, but  the critical radius $R_{cl}$ decreases. The scaling is stopped, when the
critical radius $R_{cl}\sim\xi$, i.e. when  
$l_-= ln(\frac{\xi}{a_c}) \simeq \ln(\frac{R_c}{a_c})$. 

    It is obvious that $l_- = \ln(\frac{R_c}{a_c})\rightarrow \infty$
as $u_+ \rightarrow 0$.
On the other hand the scaling constraint on the
temperature   is that $\ln\frac{\xi}{a_c} \rightarrow \infty $
as  $\epsilon\rightarrow 0$.
Setting $l = l_- \rightarrow \infty$ into the partition
function, we see that it is well defined as only if 
\begin{equation}
R_c \cong a_c(u_+)^{-\frac{1}{\lambda_3}} \cong \xi \cong a_c
\mid \epsilon \mid^{-\frac{1}{\lambda_+}}
\end{equation}
whence we find that the critical velocity
\begin{equation}
V_c \simeq V_{0} \left(1-\frac{T}{T_c}\right)^p
\label{scale-rel}
\end{equation}
where $V_{0}=\hbar/m a_c$ and  the critical exponent $p ={\frac{\lambda_3}{\lambda_+}}= 1$, that observed in Refs\cite{Varo,Ihas,Davi,Varoq,Ihas1,Tyle}.

    The proposed mechanism of the  vortex nucleation 
has a very general character. It is definitely
applicable to systems like orifices with different geometry.
The latter dictates the shape 
of the optimal vortex ring segments fluctuatively generated. 
For orifice of a square or a rectangular cross section the relevant
will be  quarter-vortex ring segments. They prefer to  be nucleated at corners.
There a vortex-ring segments "plasma"
is  spontaneously generated. Via this mechanism   the barrier for such a nucleation vanishes and the vortex (or a vortex ring) is nucleated.
The scaling relation for critical velocity will again take the derived
universal form(\ref{scale-rel}) with $p=1$.
The  vortex nucleation
is related to some kind of a phase transition, which occurs near the surface
with the critical width  $R_c$.
In the limit when $T\rightarrow T_c$ the
width $R_c$ increases and this "near surface phase transition"
 transforms into the bulk  $\lambda-$phase transition driven by
 a generation of half-vortex rings  but not a  generation of full vortex rings as
in the WS model. Here
 the coefficient  $A_0$ has also decreased.
Such a change  reduces the critical temperature but has no effect on the
critical indices. Thus, our findings offer a new driven mechanism for the
$\lambda-$ phase transition -- a generation of half-vortex rings .
Probably,  for a complete description of the $\lambda-$ phase transition
both  half-vortex rings  and 
full vortex rings must be taken into account. This may give a more reasonable value for the coefficient
$A_0$.  

There is nothing 
in the theory which restricts it to pure bulk $^4$He. 
Our  findings  give an explanation for a number of  phenomena 
in  Vycor glasses, where there are narrow
channels. Then, one has to take into account
the curvature
of the walls of these narrow channels. Therefore, instead of
the half-vortex rings for the plane geometry,
  the optimal shape of fluctuations
 will be  smaller segments
 of the vortex rings. This shape depends on the curvature, i.e. on the
radius of the narrow channel. Because of this dependence 
the coefficient $A_0$ in the scaling equations
 and the critical temperature of the phase transition  decrease while the critical indices   i.e.  the universality class of the phase transition
 remain the same.
 This    explains why in Vycor  glasses the critical
temperature decreases with the decrease of the diameter 
of the channels while the character of the $\lambda-$phase 
 transition  remains the same. Therefore the critical velocity
in the Vycor is described by the same formula as in the absence of 
the Vycor \cite{Tyle}. Thus, the theory works equally well 
for the flow of $^4$He in Vycor and
probably for Xerogel
and Aerogel glasses\cite{Tyle}. However, the fractal structure of Aerogel
glasses may give some peculiarities.

The other relevant systems are superfluid films of finite thickness.
 Ambegaokar et al\cite{Ambe} shown that there
with the temperature occurs a crossover from  $2D$  to $3D$.
With this crossover the character of the vortex nucleation changes.  
 We discuss this here only  qualitatively.
The quantative treatment must include  anisotropy effects
created by the flow as it was also indicated by perturbation theory\cite{Haus,Rem3}.
The behavior of the critical velocity in the films depends on whether
the thickness of the film $d$ is bigger ($d>R_c$) or smaller ($d<R_c$)
than the critical size  $R_c$ of the surface vorticity sheet . When the the thickness of the film
is smaller than the critical size  $R_c$ of the surface vorticity sheet, i.e. $d<R_c$, one may expect
that the BKT $2D$ V-A pair unbinding process holds\cite{Bowl}.  
In this case the critical velocity $V_c$ is estimated from the condition
that the coherence length of BKT transition at $T<T_c$ is equal to
a critical radius of the vortex pair separation, i.e. $\xi_-=r_c$.
The coherence length estimated in  Ref.\cite{Ambe} is
$\xi_-=a_{c2} \exp(b/\sqrt{(1-T/T_c)})$, while the critical radius of the vortex
pair separation is $r_c=\hbar/m V_c$ \cite{Gill}. Whence we
find that 
\begin{equation}
V_c=\frac{\hbar}{m a_{c2}} \exp(-b/\sqrt{(1-T/T_c)}).
\label{scale-rel-2}
\end{equation}
where $a_{c2}$ is a vortex core radius which depends on the thickness of the
film\cite{Saar}.
On the other hand when the film thickness $d>R_c$ the critical
velocity obeys eq.(\ref{scale-rel}) with $p=1$.
However, the width $R_c$ of the surface vorticity sheet    depends on
the temperature. The value of $R_c$ increases when $T\rightarrow T_c$.
Therefore,
 there exists the region of the temperatures
(near $T_c$)
where the value $R_c > d$ and the critical velocity in the film 
is described by
the expression (\ref{scale-rel-2}).
 With 
 decreasing temperature it may happen that the value $R_c$
becomes smaller than the value $d$ ($R_c<d$). Then the critical velocity
 is described by eq. (\ref{scale-rel}) with $p=1$.
Therefore, with  decreasing   temperature, there occurs
a crossover in the  behavior of the critical velocity of the film 
 from the regime described by eq.(\ref{scale-rel-2})
to the $3D$ regime.  Probably, to confirm this picture  additional experiments are needed ( see, also, Fig.20 in Ref.\cite{Repp}).

 Similarly  a vortex may be nucleated both in
the rotating superfluid $^3HeB$ and in superconductors. 
Indeed a  surprising  scaling relation for the
critical velocity of the  vortex nucleation in rotating $^3HeB$
has been observed\cite{Part}. The  huge vortex nucleation barrier
rules out the conventional mechanisms like tunneling and activation.
There is only one
possibility here to generate the vortex: i.e.  via a creation of a critical the surface vorticity sheet  the 
attraction of the vortex to the wall is screened, the barrier vanishes
and an intrinsic instability for the vortex creation
observed in \cite{Part} occurs.
However, the core of these $^3HeB$ vortices
 proportional to the BCS coherence length is growing to
infinity with its critical exponent when temperature rises.
 This makes a strong difference with the case considered
in the present paper although in general the proposed physical
mechanism of the vortex nucleation via fluctuative 
half-vortex rings  are also applicable here.
   The critical
exponents as well as the scaling equations may vary and probably be very
different from the obtained ones.

    In summary, we proposed the fluctuative mechanism of the vortex
nucleation, which is similar to the 
original BKT mechanism. 
The nucleation of the vortex is usually prevent by its attraction to the wall.
Loosely speaking, this attraction is due to  mirror forces
 to the mirror vortex. For the half-vortex ring  discussed this will an attraction to the other mirror part of
the vortex ring behind the wall which is a mirror image 
of this half-vortex ring.
If we consider the single half-vortex ring  , its penetration into the volume,
goes via transition through a very high barrier associated with this
"coulomb" attraction to the wall. However these  "coulomb" forces may be screened if in the neighborhood of the wall a large number of the half-vortex rings  of the small sizes will be created. These half-rings will create
a some kind of "plasma" located mostly near the wall, 
which, in turn, screens this attraction of the nucleated vortices to the wall. 
As a result of this screening the single vortex
may easily penetrate into the volume.
We have also the following main results accomplishing the proposed
mechanism: 1) In a flow near a surface a critical 
surface vorticity sheet occurs and then a spontaneous barrierless vortex nucleation
is started. It seems that a recent analysis of
 various experimental data supports this idea\cite{Part}. 
2) Because the half vortex rings have half the energy of the 
 full vortex rings of the same radius, they are more important both for 
the vortex nucleation and for $\lambda$-phase transition.
 3) The theoretical expression for the critical velocity (main result,
eq.(21)) has been obtained for the first time.  In the
 derivation the renormalization group equations have been used.
The proposed mechanism may
be further confirmed or rejected by experimental studies of the
critical exponents, which, according to the scaling hypothesis, obey
universal relations.

 I thank     Efim Brenner, Micke Fogelstr\"om, S.V. Iordanskii, J. Kurkij\"arvi, M. Saarela, G.E. Volovik and Gary Williams for  discussions and Academy of Finland for support.

*) also, Department of Physics, Loughborough University of Technology,
Loughborough, Leicestershire, LE11 3TU, UK


\begin{references}
\bibitem{Will-unb} The importance of the angular anisotropy of the flow
for the vortex unbinding in helium films was also independently
recognized by G. Williams (private communications).
\end{references}
\end{document}